\documentclass{jps-cp}
\usepackage{txfonts} 
\title{Strangeness Production in Jets and Underlying Events in pp Collisions at $\sqrt{s}$ = 13 TeV with ALICE}

\author{Pengyao \textsc{Cui} for the ALICE Collaboration$^{1}$}

\inst{$^{1}$Central China Normal University}

\email{pengyao.cui@cern.ch}

\recdate{January 17, 2019}

\abst{
The production of $\mathrm{K^0_S}$, $\Lambda$ and $\Xi$ is measured in jets and underlying events in pp collisions at $\sqrt{s}$ = 13~TeV with ALICE detectors. The corresponding $\Lambda/\mathrm{K^0_S}$ and $\Xi/\Lambda$ ratios are presented as a function of transverse momentum. The hard scatterings are tagged by charged-particle jets ,which are reconstructed with the anti-$k_\mathrm{T}$ algorithm using a resolution parameter $R$ = 0.4. The flow-like correlations and the enhanced production of multi-strange hadrons in small colliding systems may be described by soft components.
}

\kword{pp collisions; Strange particle production; Baryon to meson ratio, Jet fragmentation}

\begin{document}
\maketitle

\section{Introduction}
The transverse momentum ($p_\mathrm{T}$) dependence of the baryon-to-meson yield ratio in hadronic and nuclear collisions is sensitive to the collective expansion of the system, the partonic recombination into hadrons, the jet fragmentation and hadronization. The significant enhancement of such ratio has been observed at intermediate-$p_\mathrm{T}$ ($2 < p_\mathrm{T} < 6 $~GeV/$c$) in high multiplicity pp and p-Pb collision events with respect to lower multiplicity events~\cite{MultiDepend}. However, the origin of the enhancement still remains an open question. There has been an investigation on $\mathrm{V^0s}$ ($\mathrm{K^0_S}$ and $\Lambda$) production in jets in p-Pb collisions at $\sqrt{s_\mathrm{NN}}$ = 5.02~TeV with ALICE~\cite{V0InJet}. In this contribution, we explore the connection between the baryon-to-meson ratio enhancement and jet production via the measurement of the $p_\mathrm{T}$-differential spectrum of strange and multi-strange particles ($\mathrm{K^0_S}$,  $\Lambda$ and $\Xi$) in pp collisions at $\sqrt{s}=13$~TeV, both inclusively and within energetic jets. The results set new constraints on the particle production mechanisms in jets, and provide new insight in the understanding of the origin of flow-like correlations observed in small colliding systems.

\section{Analysis Strategy}

This analysis is based on 240 million pp collision events at $\sqrt{s}$ = 13~TeV collected by the LHC-ALICE detector~\cite{ALICEPerformance}. The spatial coordinates of the pp collisions are reconstructed with the Inner Tracking System (ITS). The event trigger information is provided by V0 and ITS detectors. 

We use charged particle jets to tag hard processes. Charged particle tracks are reconstructed using the Time Projection Chamber (TPC) detector. The tracks are selected within  the range of $|\eta| < 0.9$ and  $p_\mathrm{T} > 0.15~\mathrm{GeV}/c$. The charged particle jets are reconstructed with the anti-$k_\mathrm{T}$ algorithm and resolution parameter $R$ = 0.4.

In order to reconstruct (multi-)strange particles the following decay channels are used: $\mathrm{K^0_S} \to \pi^+ + \pi^-$, $\Lambda \to \mathrm{p} + \pi^-$ and $\Xi^- \to \Lambda + \pi^- \to \mathrm{p} + \pi^- + \pi^-$. The decay daughters are reconstructed and identified by their specific energy losses, $\mathrm{d} E / \mathrm{d} x$, in the TPC. The strange particles from inclusive events are reconstructed.

We only match strange particles with a jet when the distance within $\eta - \phi$ plane between the particles and the jet axis is smaller than the resolution parameter $R\mathrm{(V^0 (\Xi), jet)} < R$. There are particles from the Underlying Event (UE), not originated from jet fragmentation or generated from soft components in the matching region. In this work we estimate the UE contribution, which has to be subtracted, as follows:
\begin{itemize}
  \item Perpendicular Cone (PC): strange particles reconstructed in a cone perpendicular to the jet axis.
  \item Outside Cone (OC): strange particles reconstructed outside the jet cone of any of the reconstructed jets in the event.
  \item Non-jet events (NJ): strange particles found in events without any jet having $p_\mathrm{T} > 5~\mathrm{GeV}/c$.
\end{itemize}
We use the PC estimator as the default one. The OC and NJ estimators are used to evaluate the uncertainties on UE. The strange particles from the jet fragmentation (JE) are obtained after subtracting the UE background from those matched with jets.

The $p_\mathrm{T}$-differential density $\mathrm{d} \rho/ \mathrm{d} p_\mathrm{T}$ of strange particles is defined as :
\begin{equation}
  \frac{\mathrm{d} \rho}{\mathrm{d} p_\mathrm{T}} = \frac{1}{N_\mathrm{event}} \times \frac{1}{\langle \mathrm{Area} \rangle } \times \frac{\mathrm{d} N}{\mathrm{d} p_\mathrm{T}}
\end{equation} 
in which the $N_\mathrm{event}$ is the number of events, $\langle \mathrm{Area} \rangle $ is the averaged acceptance per event of strange particles ($\Delta \eta \times \Delta \varphi$) and $\frac{\mathrm{d} N}{\mathrm{d} p_\mathrm{T}}$ is the strange particle yields in every $p_\mathrm{T}$ bin.

\section{Results}

\begin{figure}[tbh]
\begin{center}
\includegraphics[width=.53\textwidth]{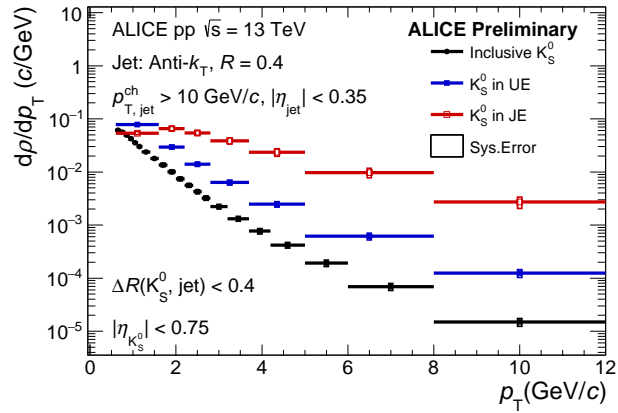}
\end{center}
\caption{(color online) The $p_\mathrm{T}$-differential density of $\mathrm{K^0_S}$ in jets with $p_\mathrm{T, jet}^\mathrm{ch} > 10~\mathrm{GeV/}c$ and underlying events in pp collisions at $\sqrt{s}$ = 13 TeV. Results are compared to the inclusive $\mathrm{K^0_S}$ $p_\mathrm{T}$ distribution.}
\label{fig:K}
\end{figure}

\begin{figure}
\begin{center}
\includegraphics[width=.53\textwidth]{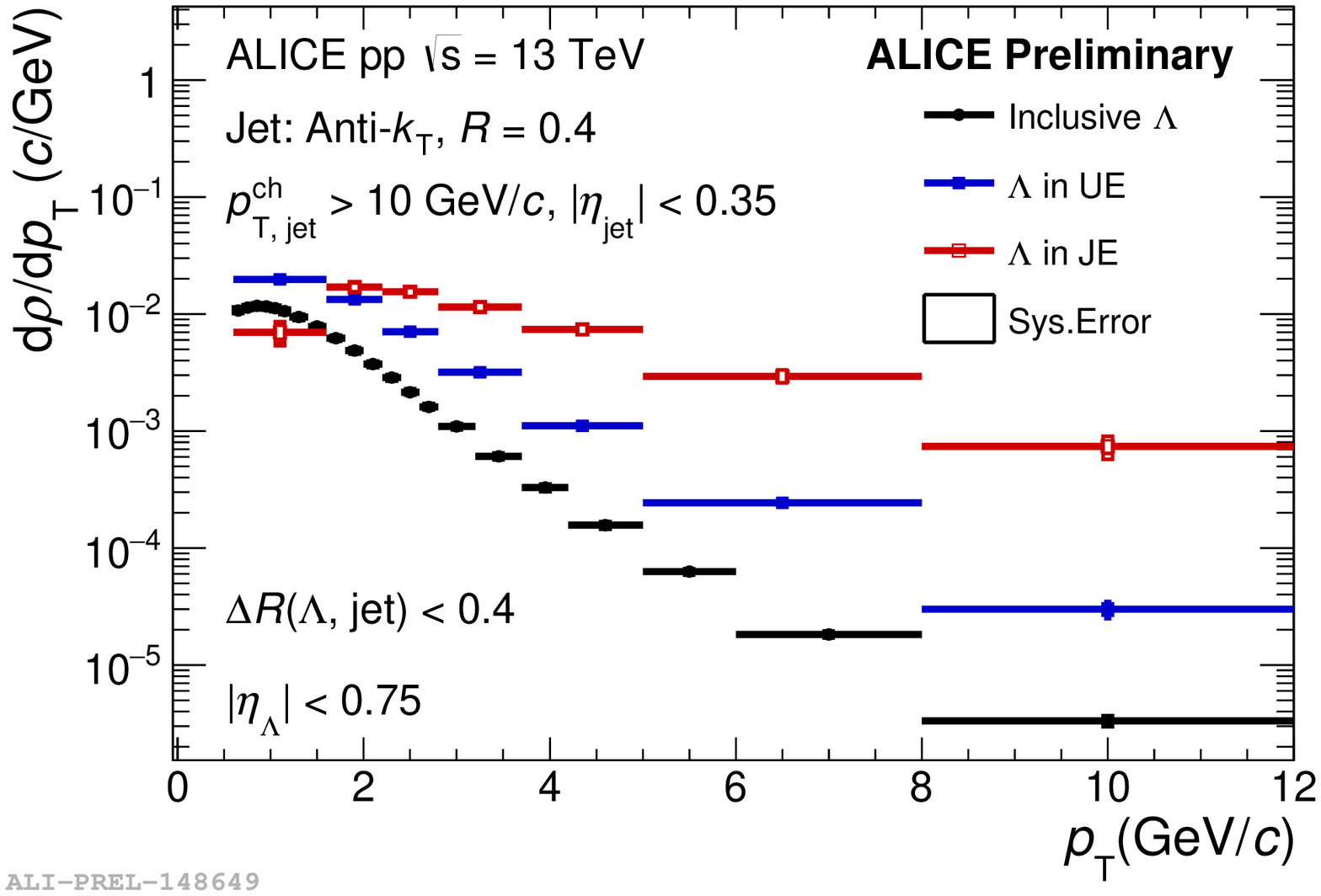}
\end{center}
\caption{(color online) The $p_\mathrm{T}$-differential density of $\Lambda$ in jets with $p_\mathrm{T, jet}^\mathrm{ch} >10~\mathrm{GeV/}c$ and underlying events in pp collisions at $\sqrt{s}$ = 13 TeV. Results are compared to the inclusive $\Lambda$ $p_\mathrm{T}$ distribution.}
\label{fig:L}
\end{figure}

\begin{figure}[tbh]
\begin{center}
\includegraphics[width=.53\textwidth]{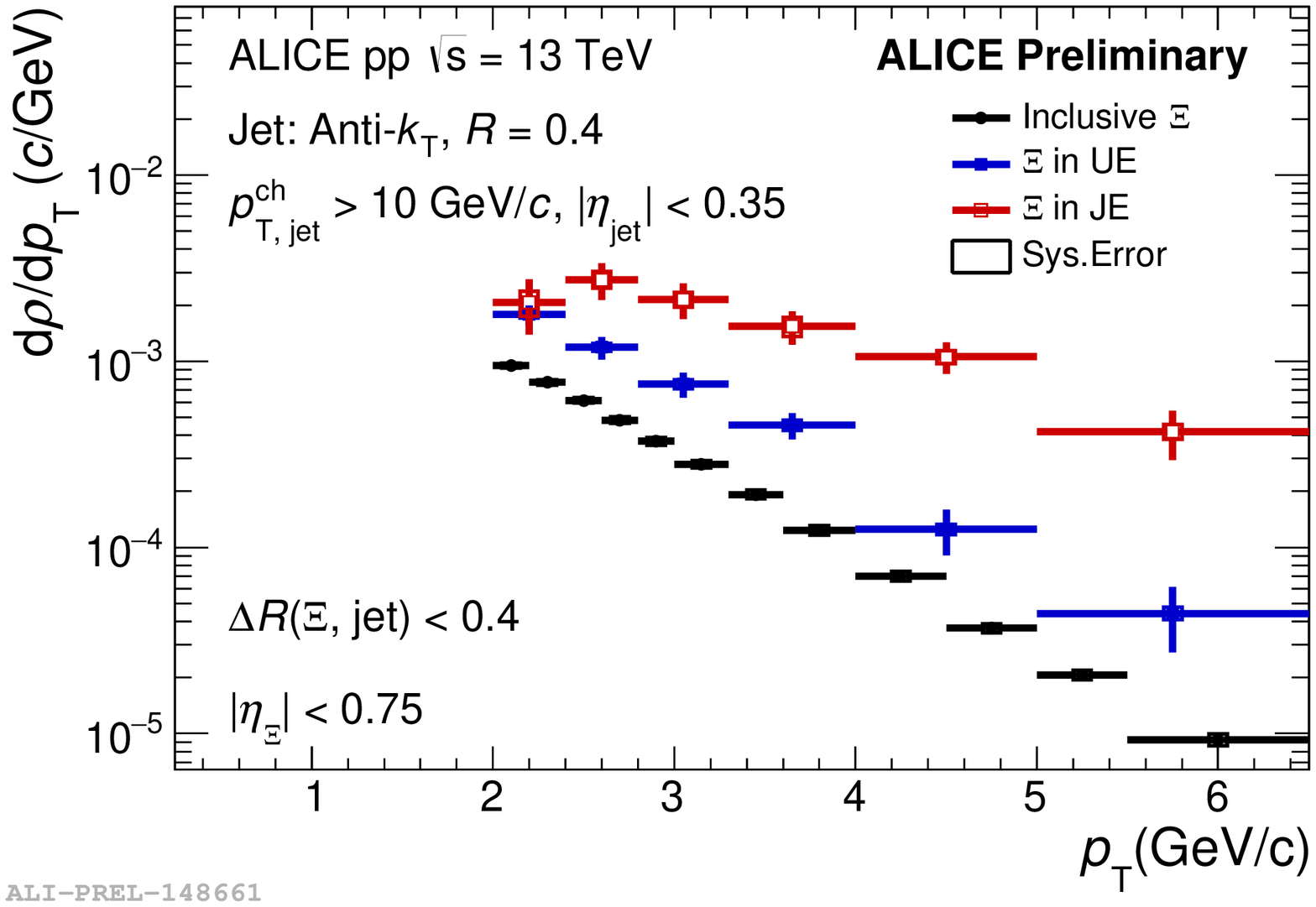}
\end{center}
\caption{(color online) The $p_\mathrm{T}$-differential density of $\Xi^- (\overline{\Xi}^+)$ in jets with $p_\mathrm{T, jet}^\mathrm{ch} >10~\mathrm{GeV/}c$ and underlying events in pp collisions at $\sqrt{s}$ = 13 TeV. Results are compared to the inclusive $\Xi^- (\overline{\Xi}^+)$ $p_\mathrm{T}$ distribution.}
\label{fig:X}
\end{figure}

Figures \ref{fig:K}-\ref{fig:X} show the $p_\mathrm{T}$-differential density of $\mathrm{K^0_S}$, $\Lambda$ and $\Xi$ spectrum in jets (red points), underlying events (blue points) and inclusive sample (black points). The spectra in jets are harder than those from the underlying event.

\begin{figure}[tbh]
\begin{center}
\includegraphics[width=.53\textwidth]{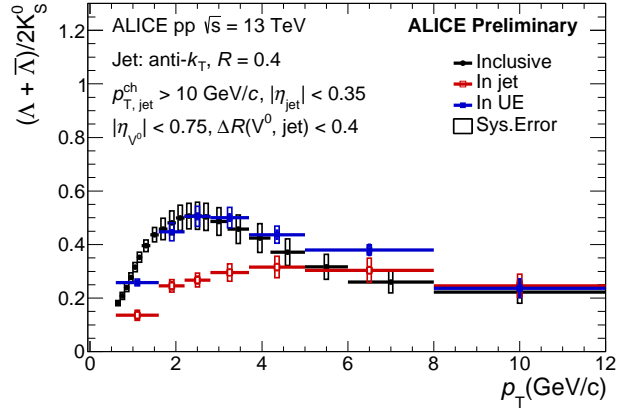}
\end{center}
\caption{(color online) The $\Lambda/\mathrm{K^0_S}$ ratio in jets with $p_\mathrm{T, jet}^\mathrm{ch} > 10~\mathrm{GeV/}c$ and underlying events in pp collisions at $\sqrt{s}$ = 13 TeV. The ratios are compared to the inclusive distribution.}
\label{fig:LK}
\end{figure}

The $\Lambda/\mathrm{K^0_S}$ ratios in jets with $p_\mathrm{T, jet}^\mathrm{ch} > 10~\mathrm{GeV}/c$ and underlying events are presented in Fig. \ref{fig:LK}. We observe that the ratio in the underlying event is the same as the inclusive event ratio. The ratio in jets is significantly smaller than the ratio in UE (or inclusive event) at low and intermediate $p_\mathrm{T}$ ($2 < p_\mathrm{T} < 4~\mathrm{GeV}/c$). 

\begin{figure}
\begin{center}
\includegraphics[width=.53\textwidth]{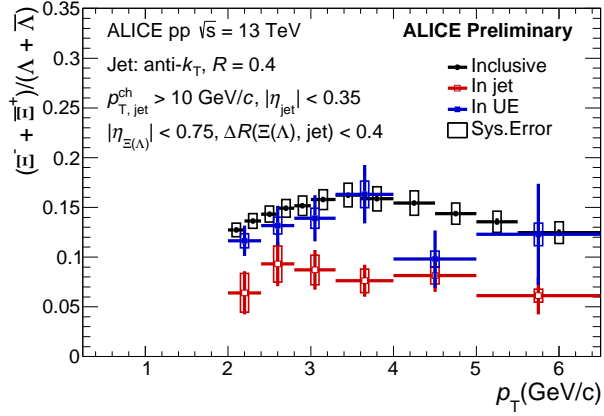}
\end{center}
\caption{(color online) The $\Xi/\Lambda$ ratio in jets with $p_\mathrm{T, jet}^\mathrm{ch} > 10~\mathrm{GeV/}c$ and underlying events in pp collisions at $\sqrt{s}$ = 13 TeV. The ratios are compared to the inclusive distribution.}
\label{fig:XL}
\end{figure}

Figure \ref{fig:XL} shows the $\Xi/\Lambda$ (multi-strange particle to strange particle for ``baryons") ratio in jets and underlying events. As observed for the $\Lambda/\mathrm{K^0_S}$ ratio, the $\Xi/\Lambda$ ratio in the underlying event is the same as inclusive one and has a bump at intermediate $p_\mathrm{T}$. However, the ratio in jets is different with respect to the inclusive (or UE) ratio, and is not strongly dependent on $p_\mathrm{T}$.

\section{Conclusions}

 The spectra of $\mathrm{K^0_S}$, $\Lambda$ and $\Xi$ in jets are always harder than that in UE. This may hint that the baryon and meson production mechanism is the same. The enhancement of $\Lambda/\mathrm{K^0_S}$ ratio is not present when the particles are within an energetic jet in pp collisions. The enhancement of baryon to meson ratio in high multiplicity events may be attributed to soft components of the collisions. The $\Xi/\Lambda$ ratio in jets is independent on $p_\mathrm{T}$. The enhanced production of multi-strange hadrons in high-multiplicity events~\cite{StrangeEnhance} may also be explained by soft components of the collisions.

\section*{ACKNOWLEDGMENT}
This work is supported by the National Key Research and Development Program of China (2016YFE0100900) and the National Science Foundation of China (Grant No. 11875143 and 11805079).


\end{document}